\documentclass[prl,twocolumn,superscriptaddress,showpacs,preprintnumbers,amsmath,amssymb,floatfix]{revtex4}
\usepackage[colorlinks=true, linkcolor=blue, citecolor=blue]{hyperref}
\usepackage{epsfig}

\usepackage{color}
\newcommand{\be}{\begin{equation}}
\newcommand{\ee}{\end{equation}}
\newcommand{\ba}{\begin{array}{rcl}}
\newcommand{\ea}{\end{array}}
\newcommand{\ds}{\displaystyle}
\newcommand{\md}[1]{\left|#1\right|}

\begin{document}

\title{Markov State Modeling of Sliding Friction}

\author{F. Pellegrini}
\affiliation{SISSA, Via Bonomea 265, I-34136 Trieste, Italy}
\affiliation{CNR-IOM Democritos National Simulation Center, Via Bonomea 265, I-34136
Trieste, Italy}
\author{F.P. Landes}
\affiliation{International Centre for Theoretical Physics (ICTP), Strada Costiera 11,
I-34151 Trieste, Italy}
\author{A. Laio}
\affiliation{SISSA, Via Bonomea 265, I-34136 Trieste, Italy}
\author{S. Prestipino}
\affiliation{Universit\`a degli Studi di Messina, Dip. di Scienze Mat. ed Inf.,
di Scienze Fis. e di Scienze della Terra, Contrada Papardo, I-98166 Messina, Italy}
\affiliation{CNR-IPCF, Viale F. Stagno d'Alcontres 37, I-98158 Messina, Italy}
\author{E. Tosatti}
\affiliation{SISSA, Via Bonomea 265, I-34136 Trieste, Italy}
\affiliation{CNR-IOM Democritos National Simulation Center, Via Bonomea 265, I-34136
Trieste, Italy}
\affiliation{International Centre for Theoretical Physics (ICTP), Strada Costiera 11,
I-34151 Trieste, Italy}

\date{\today}
\begin{abstract}
Markov State Modeling has recently emerged as a key technique for analyzing rare events in thermal equilibrium molecular simulations 
 and finding metastable states. 
Here we export this technique 
to the study of friction, where strongly non-equilibrium events are induced 
by an external force.
The approach is benchmarked on 
the well-studied Frenkel-Kontorova model, 
where we demonstrate the unprejudiced identification of the minimal basis 
microscopic states necessary 
for describing sliding, stick-slip and dissipation. 
The steps necessary for the application to realistic frictional systems are highlighted.
\end{abstract}


\maketitle

Despite the relevance of friction 
between solids 
from the macroscale to the nanoscale, 
its physical description still needs 
theoretical basis and understanding. 
Even the simplest, classical atomistic sliding problem has too many degrees of freedom, and
there is so far no 
method for the unprejudiced 
identification of a few dynamical 
collective
variables
suitable for a mesoscopic description 
of fundamental sliding events such as stick-slip~\cite{gnecco_meyer2015}.
In the 
field of  
equilibrium 
biomolecular simulations, where computational
scientists often meet similar problems, powerful tools have been 
developed in the last decade, aimed at identifying the
relevant metastable conformations, the reactions paths, and  
the rates associated to transition events between them. 
In particular, Markov State Models~\cite{Schwantes2014, Noe2013, Bowman2014, Schutte2015}
(MSMs) have emerged as a key technique, with clear
theoretical foundations and great flexibility. 
In that approach, the dynamical trajectory in phase space of 
a large collection of molecular entities
is projected onto a much smaller space defined by a discrete
set of states that are deemed typical, and the dynamics is reduced
to Markovian jumps between these states. In most cases so far 
MSMs were applied to systems at equilibrium, where a stationary 
measure is defined and the Markov description is natural.
In the physics of friction we deal with strongly nonequilibrium 
dynamics, even in steady state sliding.
 Application of MSMs to nonequilibrium problems is still in its infancy,
with apparently only one instance, related to periodic driving~\cite{Wang2015}.

Here we show how the MSM framework can be extended to the
study of nanofriction dynamics.
To demonstrate that concretely, we choose one of the simplest tribological models,
the one-dimensional Frenkel-Kontorova (FK) model~\cite{Frenkel1938}
in its atomic stick-slip regime~\cite{Braun1998, Paliy1997}. 
The MSM construction leads to the identification of a handful of natural variables  
which describe the steady-state dynamics  
of friction in this model.

Starting from the set of configurations obtained with a simulation of steady-state sliding,
the first step of the construction is to define 
a metric in the high dimensional phase space of the original model,
then used to identify a small number of 
{\em microstates}, by means of a recently proposed 
clustering algorithm~\cite{Rodriguez2014}. 
The statistics of transitions between configurations
is shown to be compatible with a description as a Markov process between the microstates.
The highest eigenvectors of the transfer operator provide 
a novel characterization of the slowest modes of frictional motion.
The space of microstates can be further 
coarse-grained into a few {\em macrostates}
using standard coarse-graining methods, ~\cite{Deuflhard2004, Weber2005} 
finally yielding a compact
MSM description. 
In it, the time evolution of observables such as  frictional work and  displacement 
still reproduces the  main features of the original frictional dynamics.
The states of this Markov process reveal the definition of the collective variables describing friction, 
which for the simple FK model are the kink-antikink populations, 
but should be naturally found also in out of equilibrium 
sliding systems of higher and generic complexity.

\textit{The transfer operator and matrix} --- 
Our analysis is based on the
Transfer Operator (TO) formalism~\cite{Bowman2014}. 
Denote by $\Pi^{\tau}(X\rightarrow X')$ the probability to go from a configuration 
$X^{t}=X$ at time $t$ to $X^{t+\tau}=X'$ at time $t+\tau$.
While 
$\Pi^{\tau}$ is a continuous
process
and would take infinite time to sample, 
we build a coarse-grained TO by partitioning the configuration space 
into microstates (ensembles of similar configurations) $\{c_{\alpha}, \alpha=1,\dots, n_c\}$. 
Between these microstates the restricted TO is
a finite $n_c\times n_c$ Transfer Matrix (TM) with the generic element
 $\Pi_{\alpha\beta}^{\tau}=\int_{X\in c_{\alpha}}\int_{X'\in c_{\beta}}{\rm d}X{\rm d}X'P(X)\Pi^{\tau}(X\rightarrow X')$,
the probability to go 
from $c_\alpha$ to $c_\beta$ in time $\tau$.
This TM contains less detail than the full
 $\Pi_{\alpha\beta}^\tau$. 
Being simpler, it is more informative,
and can be sampled with satisfactory statistics in much less time. Of course $\Pi_{\alpha\beta}^{\tau}$ also depends on the lagtime $\tau$, 
but there are techniques to control the error related to the choice of this parameter 
(see SI 1.1).
Given the TM, we calculate its eigenvalues $\{\lambda_{i}\}$ and left eigenvectors
$\{\vec{\chi}_{i}\}$. 
Because we are not in equilibrium, detailed balance does not hold, 
the TM is not symmetric and
the eigenvalues are not necessarily real. However,
they still satisfy $\md{\lambda_{i}}\le1$ by the Perron-Frobenius
theorem. The largest (modulus-wise) eigenvalue is 
exactly $1$, and if the evolution is ergodic there is only one such eigenvalue.
The eigenvector $\vec{\chi}_{1}$   
represents the invariant, steady state distribution, endowed with nonzero sliding current.
The eigenvectors $\vec{\chi}_{i}$ with $\md{\lambda_{i}}\simeq 1$ form
the so-called Perron Cluster~\cite{Deuflhard2004}. They characterize the long-lived excitations of the steady state,
which decay with 
long characteristic times $\tau_{i}=-\tau/\ln\md{\lambda_{i}} \gg \tau$, while
oscillating with period $\tau/\arctan({\rm Im}\lambda_{i}/{\rm Re}\lambda_{i})$.

\begin{figure}
\centering 
\includegraphics[width=0.48\textwidth]{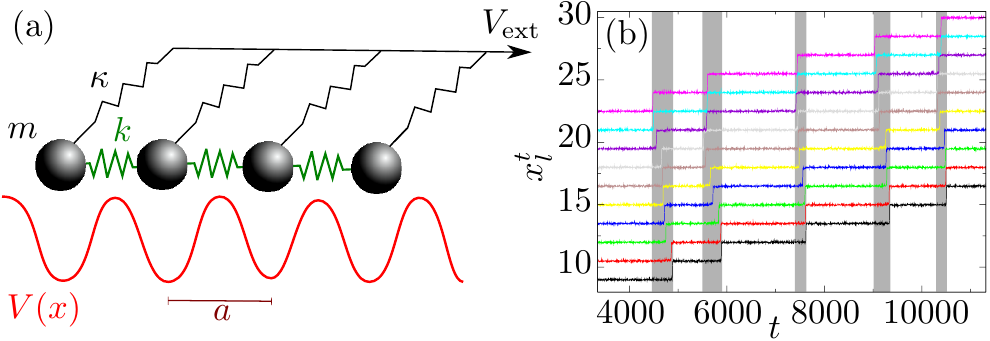} 
\caption{
\label{evol}
(a) Schematic of the FK system.
(b) Sample of steady-state motion of $L=10$ particles  
with parameters $k=0.04$, $A=0.1$, $a=1.5$, $m=1$, $\gamma=1$,
$\beta=500$, $\kappa=0.01$, $v_{{\rm ext}}=0.001$. 
The white and grey backgrounds represent 
stick and slip time domains respectively. 
}
\end{figure}

\textit{The Frenkel-Kontorova model} --- 
The one-dimensional FK model,  Fig.~\ref{evol}(a), our test case,
consists of a chain of particles dragged over a sinusoidal potential $V(x)=A\cos(2\pi x/a)$. 
Nearest neighbor springs of stiffness  $k$ link $L$ classical particles of mass $m$ and positions
$x_{i}$ whose spacing $a$ is 
commensurate with the potential. Each particles is dragged by a spring of constant $\kappa$ moving 
with constant velocity $v_{{\rm ext}}$. 
Particle motion obeys an overdamped Langevin dynamics (large damping $\gamma$), 
in a bath of inverse temperature $\beta=1/k_{B}T$:
\be\ba
x_l^{t+dt}&=&\ds x_l^t+
\left(\frac{2\pi A}{\gamma m a}\sin\left(\frac{2\pi x_l^t}{a}\right)
+\frac{\kappa}{\gamma m}(v_{\rm ext}t-x_l^t)+\right.\\
&&\ds\left.-\frac{k}{\gamma m}(2x_l^t-x_{l-1}^t-x_{l+1}^t)
\right) dt
+\sqrt\frac{2 dt}{\gamma m \beta}f^t,
\ea\ee
where $f^{t}$ is an uncorrelated Gaussian distribution 
and $dt$ is the elementary time step (here $dt=10^{-2}$).
Our input is the steady-state trajectory
of the chain, obtained by integrating these equations, mostly for the simplest case of $L=10$ (but also $L=15, 20$) 
and a sufficient duration of $10^6$ time units.

As is well known~\cite{Braun1998},
in a wide range of parameters the chain sliding alternates
long sticking periods during which particles are close to their respective
potential minima with fast slips during which 
one or more lattice spacings are gained. 
This kind of atomic stick-slip motion is 
well established for, e.g.,  the sliding of an Atomic Force Microscope  tip on a crystal surface~\cite{vanossi2013}.
The slip event involves
the formation of kink/antikink defects (large deviations
of the interparticle distance from the equilibrium value) that propagate 
along the chain and enable the global movement. A sample of steady-state 
sliding evolution can be seen in Fig.~\ref{evol}(b), 
showing the finer details of each particle's motion for a few  slip events.

With this trajectory in the $L$-particle phase space the building 
of our Markov State Model (MSM) (see e.g.~\cite{Schutte2015})
involves three steps : choice of a metric, clustering into microstates, and construction
of macrostates and their reduced TM dynamics.

\textit{Metric} ---
The phase space explored under steady sliding grows linearly with time and is thus generally 
very poorly sampled. Internal variables are free of this problem; a viable metric in phase space  
can therefore include e.g., the bond lengths  $b_{l}^{t}=(x_{l+1}^{t}-x_{l}^{t}-a)/a$.
On the other hand the inclusion of growing degrees of freedom like the 
position of the center of mass (CM): $x_{\rm CM}^{t}=\frac{1}{Na}\sum_{l=1}^{L}x_{l}^{t}$
cannot be implemented without caution.
Sampling can be improved if distinct parts of the steady-state evolution can be 
considered as equivalent.
Ideally, we could consider a portion of the evolution long enough that all relevant 
events (here, frictional slips) have occurred, 
then set ``absorbing'' boundary conditions 
for any such transition from and to the outside of this range, then averaging over many such equivalent stretches.
In the alternative approach which we adopt here, we substitute 
the absorbing
conditions with artificially periodic boundary conditions, a choice which
provides a more compelling picture 
of steady-state sliding, and where
the error involved in the transition rates can be reduced at will by extending the portion size.
In the FK system, we exploit the substrate periodicity and 
$x_{\rm CM}$ is  taken 
modulo $na$ for a chosen integer $n>1$. Under slow driving, $n=2$ is sufficient
for a correct description of slips by $a$ (atomic slip), and states divide into even and odd $x_{\rm CM}$.
If slips of $2a$, $3a$ or more became more frequent, we would simply choose a larger $n$.
The full set of steady state sliding data is used to generate many independent configurations, all treated in the same manner. 
Summing up, the metric we adopt defines the distance between configurations at times $s$ and $t$ as
\be
d_{st}=\left[(x^s_{\rm CM}-x^t_{\rm CM})_{\text{mod }2}\right]^2+
\sum_{l=1}^{L-1}\left(b^s_l-b^t_l\right)^2.
\label{Eq:dist_dB}
\ee

\textit{Microstates} --- In the second step, configurations whose relative distance is small are collected together, in $n_c$
microstates. Microstates are built by the Density Peak algorithm~\cite{Rodriguez2014}, 
which efficiently traces them as maxima of the probability density in phase space.
Given a distance $d_{st}$ between two configurations
$X^{s}$ and $X^{t}$ we estimate the local density $\rho^{s}$ in
$X^{s}$ by counting the number of configurations within a 
cutoff $d_{c}$, $\rho^s=\sum_t\theta(d_c-d_{st})$ where $\theta$
is the step function.  
One then computes the distance $\delta^{s}$ between $X^{s}$ and the closest 
configuration of higher density, $\delta^s=\min_{\rho^t>\rho^s}d_{st}$
and identifies the microstate centers as the $n_c$ points with the highest product $\delta^{s}\rho^{s}$. 
All remaining points are assigned to the microstate of highest local density. 
This clustering technique allows finding microstates of variable volume in phase space, and
well-defined cluster centers (configurations often visited), both desirable features
in building a MSM. 
The next step is the dynamics between microstates which we describe in the FK model.

\begin{figure}
\centering \includegraphics[width=0.48\textwidth]{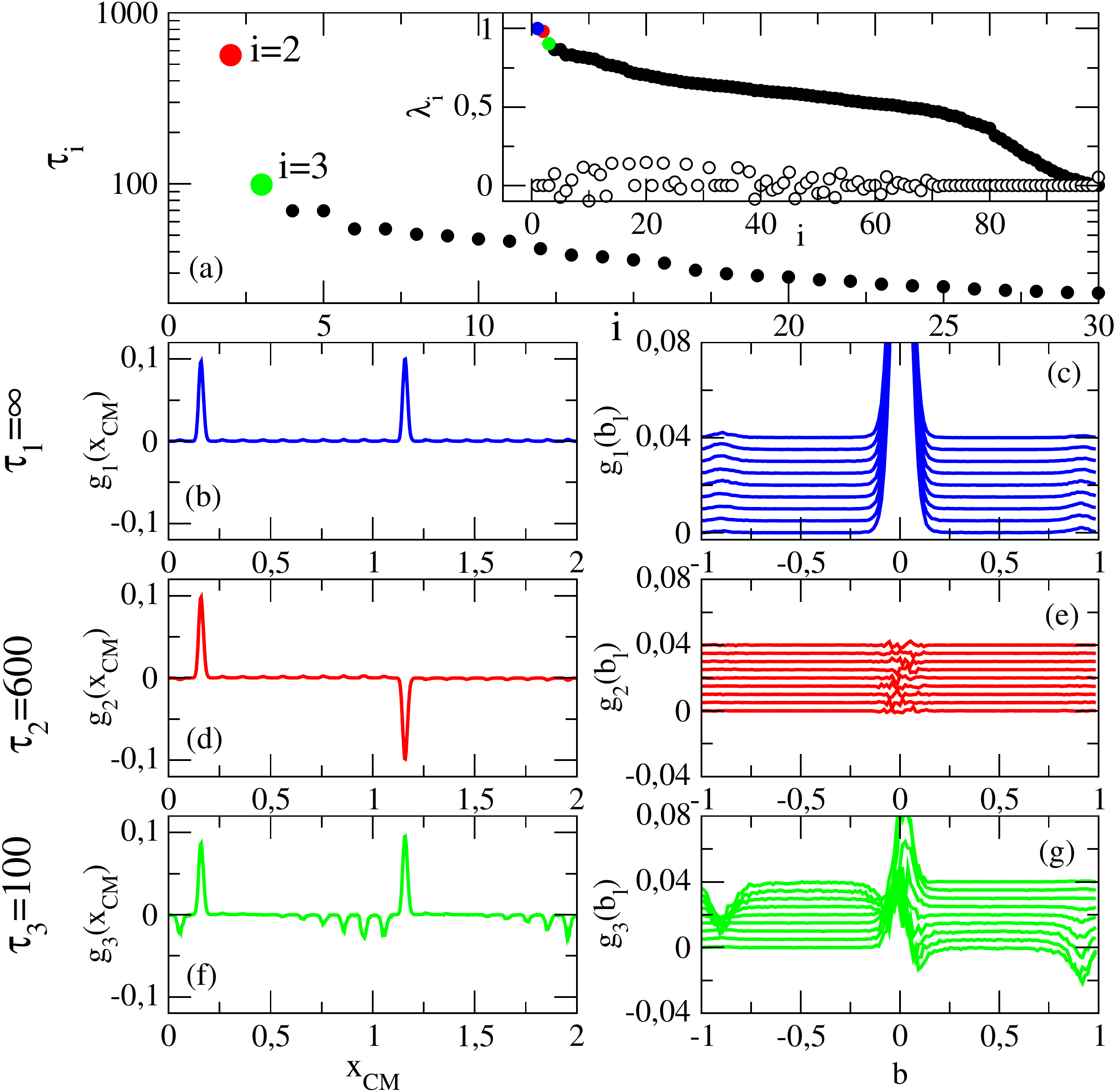} 
\caption{\label{Measure1}(a) Characteristic timescales and eigenvalues
(in the inset, imaginary part in white) of the TM 
(averaged over 10 realizations with $N=10^5$ each).
(b),(d),(f) Probability distribution $g_1(x_{\rm CM})$
and perturbations $g_i(x_{\rm CM})$ of
the center of mass position $x_{\rm CM}$ 
for the first $3$ eigenvectors of the TM.
(c),(e),(g) Probability distribution $g_1(b_{l})$
and perturbations $g_i(b_{l})$ of each bond
$b_{l}$ for the first $3$ eigenvectors of the TM 
(successive bonds $b_{l}$ are spaced vertically of $0.005$ 
each, for clarity).}
\end{figure}

We use samples of $N=10^5$ configurations (separated by the lagtime $\tau$) and 
cluster them using the metric~(\ref{Eq:dist_dB}). 
The optimal lagtime $\tau$ is determined by studying the evolution of the spectrum of the clustered TM with $\tau$ (see SI sec.~1). 
We find a plateau around the value  $\tau=10=1000 dt$, 
showing that 
the dynamics is Markovian in this range. With $\tau=10$, 
the algorithm detects a PC of $n_c\simeq 100 $ microstates.
Besides  $\lambda_1 = 1$, the spectrum of the $n_c \times n_c$ TM is characterized by a second eigenvalue $\lambda_2$ 
(see Fig.~\ref{Measure1}), corresponding to a relaxation time of $\simeq 600$, 
separated by a gap from other eigenvalues with shorter relaxation times.
The 
significance 
of the eigenmodes $\chi_i$ is clarified by considering
the probability distribution $P\left(O,t\right)$ 
of an observable $O$ at time $t$, starting from a system prepared in the mixed state $P^0_\alpha$ (probability vector to be in $c_\alpha$ at $t=0$). We have:
\be
P\left(O,t\right)=P^{\rm ss}\left(O\right)+\sum_{i>1}f_{i}g_{i}\left(O\right)e^{-t/\tau_{i}},
\ee
where $f_i=\sum_\alpha\chi_i^\alpha P^0_\alpha/P^{\rm ss}_\alpha$ 
accounts for the initial condition and 
\be
g_{i}\left(O\right)=\sum_{\alpha}\chi_i^\alpha P(O | \alpha),
\ee
where $P(O | \alpha)$ is the probability distribution of $O$ in microstate $\alpha$,
$P^{\rm ss}(O)=g_1(O)$ the steady state distribution of $O$, and $P_\alpha^{\rm ss}$ the steady state probability to visit microstate $\alpha$.
The $g_{i}(O)$ for $i>1$ represent ``perturbations'' of $P^{\rm ss}(O)$, each decaying
within the lifetime $\tau_{i}$.

In Fig.~\ref{Measure1}(b)(d)(f) we plot $g_i(x_{\rm CM})$: 
the steady state $\chi_{1}$ consists of one large
peak per period plus $9$ smaller peaks, corresponding to the
relaxed chain state and defect combinations, respectively. The
second eigenvector $\chi_{2}$ presents exactly the same features,
except for a factor $-1$ in the second period:  
in the combinations $\chi_{1}\pm\chi_{2}$  the chain CM sticks either in an odd or even position. 
The second eigenvector is thus representative of the main advancing motion
of the chain, namely the slip. Indeed $t_{2}= -\tau /\log(\lambda_2) \simeq600$ is
about half the sticking time (see Fig.~1).
In Fig.~\ref{Measure1}(c)(e)(g) we plot 
$g_i$ ($i=1,2,3$) for the bond lengths $b_l$.  
In the steady state $\chi_{1}$ each bond length
has high peaks around its value at rest  $(0)$
and smaller ones around $b\simeq\pm0.9$, reflecting the infrequent  appearance of excitations, that are kinks or antikinks. 
The second mode $\chi_2$ shows a flat distribution, since all the difference with
$\chi_1$ lies in the 
CM degree of freedom, $x_{\rm CM}$.
In fact $\chi_{3}$ displays small central peaks and more pronounced lateral
ones, corresponding to the creation (destruction for negative peaks)
of a kink or antikink (depending on the sign of $b_{l}$)~\cite{Braun1998},
excitations with shorter lifetimes. Indeed, $t_{3}\simeq100$ is 
comparable to the half-lifetime of kinks and anti-kinks.
Furthermore, we can see how
the peaks tend to be positive for the first $b_{l}$'s and negative
for the last ones, implying that the chain tends to be elongated in
its head and compressed in its tail. This shows that kink-antikink
pairs are more likely to be formed in the center of the chain,
intrinsic to the slip mechanism for this system. 
At this stage one can already identify the kink and antikink populations as the relevant collective variables of sliding, together with $x_{\rm CM}$.
While these gross features of a commensurate FK stick-slip are therefore already 
contained in the first few long-lived microstates with largest eigenvalues, 
a more accurate description must involve a quantitative analysis of the whole Perron cluster.

\begin{figure} 
\includegraphics[width=0.48\textwidth]{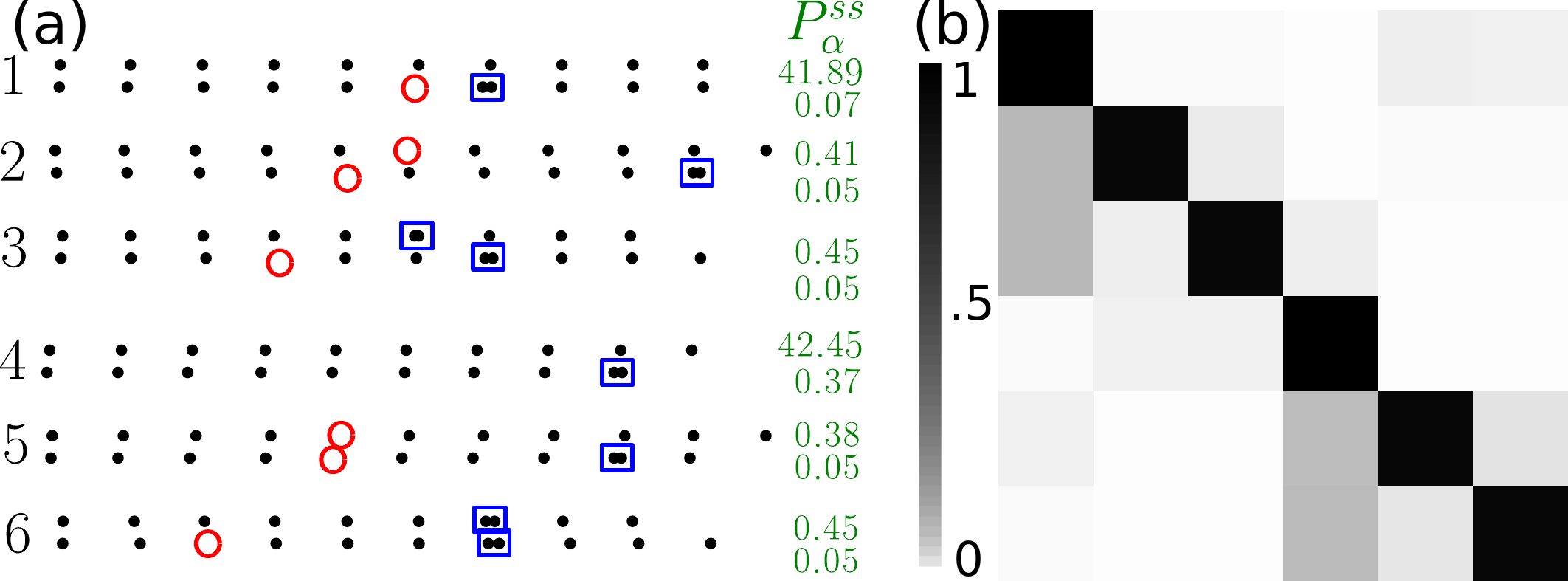}
\caption{ \textbf{(a)} 
Selection of microstates inside the six macrostates identified with PCCA+: the atoms positions relative to potential minima (black dots) 
display kinks (red circles) and anti-kinks (blue squares).
For clarity, the probability $P_\alpha^{\rm ss}$ is multiplied by $100$, and we show only $12$ of the $\sim 100$ microstates.
\textbf{(b)} Representation of the reduced transition matrix $\widetilde{\Pi}_{\alpha\beta}$ (grey scale proportional to magnitude, see text).
}\label{fig:pcca_lumped}
\end{figure}

\textit{Macrostates} --- 
In the third and final step, the $n_c$  microstates are 
coarse-grained and 
grouped into macrostates. A well established approach for that 
is the (Robust) Perron Cluster Cluster Analysis (PCCA+)~\cite{Deuflhard2004, Weber2005} (see also SI 2.1 and~\cite{CodeLandesPellegrini2016}).
Assuming for simplicity to forget the non-equilibrium breaking of detailed balance (thus building a symmetrical approximate TM, see SI 2.1 
and~\cite{Roeblitz2008,Conrad2015} for refinements), 
we find that relevant macrostates can be reduced from  $n_c\sim100$ down to as little as $\tilde n_c=6$. 
Moreover, whereas $n_c$ grows with system size $L$, 
$\tilde n_c=6$ is 
much more 
stable against $L$: we find a consistent description of the system with $\tilde n_c=6$ also for $L=15$ and $L=20$,
and the detection of the optimal $\tilde n_c$ robustly yields $\tilde n_c\in[5,9]$ (see SI, Figs.~3, 4).
In Fig.~\ref{fig:pcca_lumped} we present the six macrostates $\{\tilde{c}_\alpha\}$, displaying 
some of the microstates which they contain.

Macrostates $\tilde{c}_1$ and $\tilde{c}_4$ include the relaxed chain microstates, along with some single excitations at the tips;
$\tilde{c}_2$ and $\tilde{c}_5$ contain mostly single kinks,
while $\tilde{c}_3$ and $\tilde{c}_6$ contain mostly anti-kinks.
The microstates with (kink, anti-kink) pairs are spread between groups, with 
neighboring pairs belonging to $\tilde{c}_{1,4}$ and extended pairs to others.
The only difference between the triplets of $\tilde{c}_{1,2,3}$ and $\tilde{c}_{4,5,6}$ is in 
the value of $x_{\rm CM}$, respectively $x_{\rm CM}\approx0.15$ and $1.15$.
Overall, this description provides a qualitative understanding of the basic 
mechanisms of slips complementary to that of our kinetic analysis, and 
allows to directly read the kink and antikink populations as the collective variables describing sliding.
The $6\times6$ TM $\widetilde{\Pi}_{\alpha\beta}$ (Fig.~\ref{fig:pcca_lumped}(b)) shows,
e.g., that motion (looping through states) occurs only through excited states $\tilde{c}_{2,3;5,6}$. 
Additional details about the role of each macrostate is given in the SI sec.~2.3.

\begin{figure} 
\includegraphics[width=0.48\textwidth]{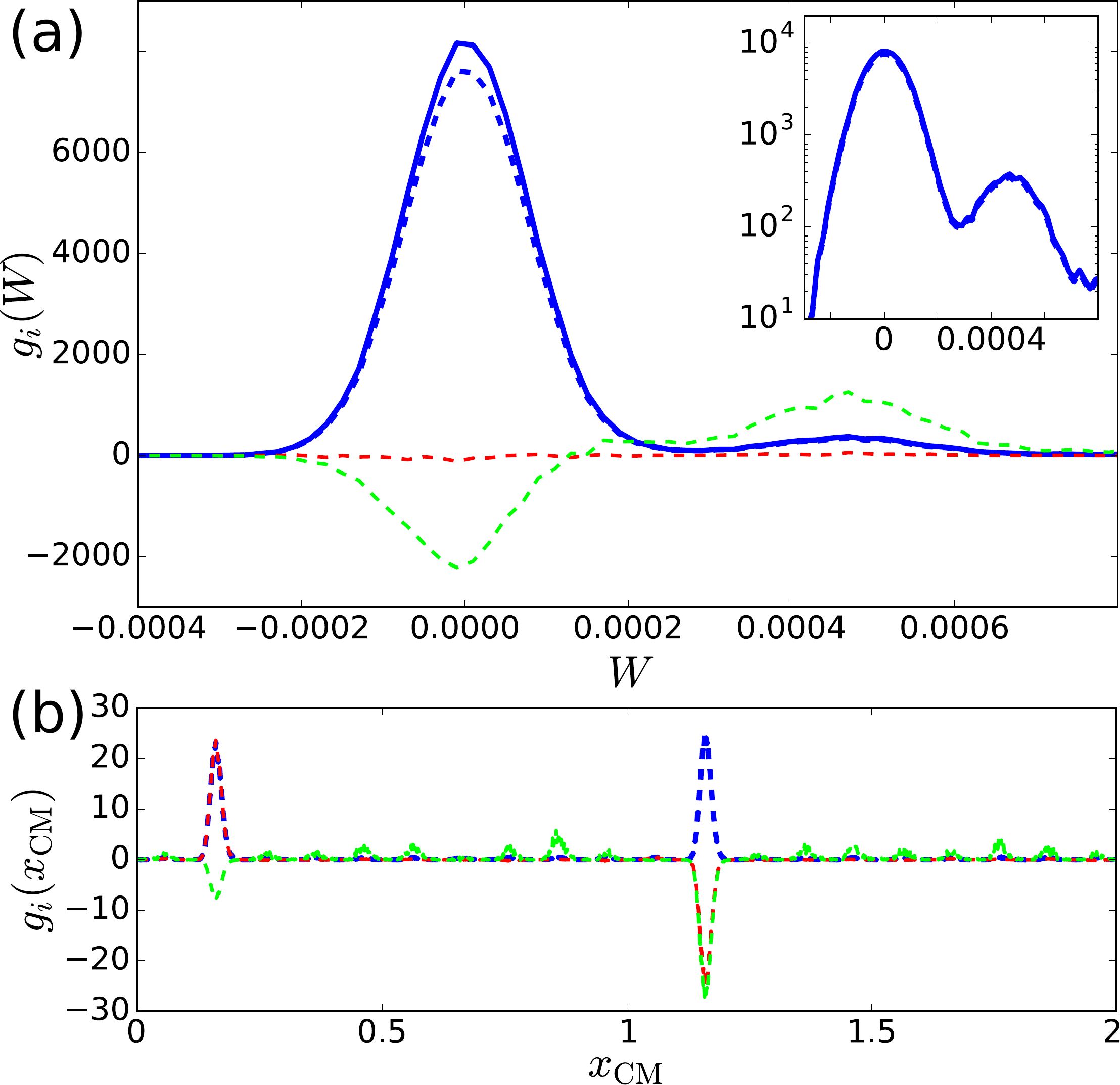}
\caption{\label{fig:work_distro}
Comparison of  \textbf{(a)} work $g_i(W)$ and \textbf{(b)} center of mass $ g_i(x_{\rm CM})$ distributions for $i=1,2,3$ (blue, red, green).
Solid lines: raw data; 
dashed lines: $\tilde n_c=6$ macrostate results.
Inset: blowup of $g_i(W)$, highlighting  the excess probability for $W>0$, signaling the positive frictional work. 
Note how $i=3$ (green) is the excitation of the steady state (blue) that populates the  $W>0$ tail. 
In \textbf{(b)}, the $i=2$ (red) excitation simply shifts  the chain by $a$.}
\end{figure}

\textit{Macrostate evolution, benchmarking} ---
In the macrostate representation, 
the probabilities
 $P_\alpha^t=P(X^t\in\tilde{c}_\alpha)$ evolve in time as
  $P_\alpha^{t+\tau} = \widetilde{\Pi}_{\alpha\beta}P_\beta^t$.
For the whole construction to be satisfactory, this coarse-grained evolution should reproduce the quantitative aspects 
of the frictional dynamics 
of the raw data (before clustering). 
To this effect, we compare the work distribution $P^{\rm ss}(W)$ in the raw data with that of the steady state 
relative to $\widetilde{\Pi}_{\alpha\beta}$, where each $\tilde{c}_\alpha$ is associated to the distribution $P(O | \alpha)$ 
within the configurations belonging to $\tilde{c}_\alpha$.
Results in Fig.~\ref{fig:work_distro}(a) confirm that the stationary work distribution in the reduced model matches well
the raw distribution. The particle current in the reduced basis is $\langle J \rangle =6.82\cdot10^{-4}$, 
compared with the exact $\langle J \rangle = 6.66\cdot10^{-4}= v_{\rm ext} / a$.
A similar agreement is found for the center of mass ($x_{\rm CM}$) 
dynamics and other observables. The steady state and
excitation modes $g_i(x_{\rm CM})$ reproduce well those of the $n_c$ 
states description (see Fig.~\ref{fig:work_distro}(b) and SI Fig.~4).
The lifetimes corresponding to these modes, and more precisely the decay of the correlation 
functions of various observables also match well the respective correlation functions evaluated on the raw data.

\textit{Conclusions} --- 
We formulated and carried out the
first analysis of frictional sliding conducted through the MSM 
method extended to non-equilibrium. After an initial choice of metric for the phase space, the approach builds in an 
unbiased manner a 
still large but
limited number of microstates 
that allow to track the effective dynamical variables of a sliding problem. 
Using
the standard PCCA+ approach, one then derives a reduced dynamics in only a handful of  macrostates.
In the chosen FK model implementation the method 
works 
well, and the coarse-graining is sharp enough to capture not only overall steady-state 
observables such as average dissipated power or average current, but also their modes of excitations and their correlations, 
as shown by the excitations $g_i(O)$. 
In this way all the important, slow dynamical features can be brought under control in a manner which is, as far as we know, 
unprecedented for violent, nonlinear frictional motion.
Further developments to efficiently improve the statistical quality could introduce a biased sampling favouring the 
exploration of rare transition event states in cases where a long ``ergodic'' trajectory cannot be generated~\cite{Chodera2007}. 
This work opens a route towards a quantitative approach to frictional dynamics, in nanoscale sliding as well as in other driven systems. 

\textit{Acknowledgments} --- 
 Work carried out under ERC Advanced Research Grant N.~320796 -- MODPHYSFRICT.  FL thanks R. Dandekar for useful discussions.

\pagebreak
\widetext
\begin{center}
\textbf{\large Supplementary material to Markov State Modeling of Sliding Friction}
\end{center}

\section{1.1 Lagtime estimation}

As usual in MSM models, we need to determine the range of lagtimes that are long enough to obtain markovianity, and thus a correct estimation of the dominant timescales, while still being shorter than the time scale of the most rapid events we want to describe.

\begin{figure*}[h!]
\includegraphics[width=\textwidth]{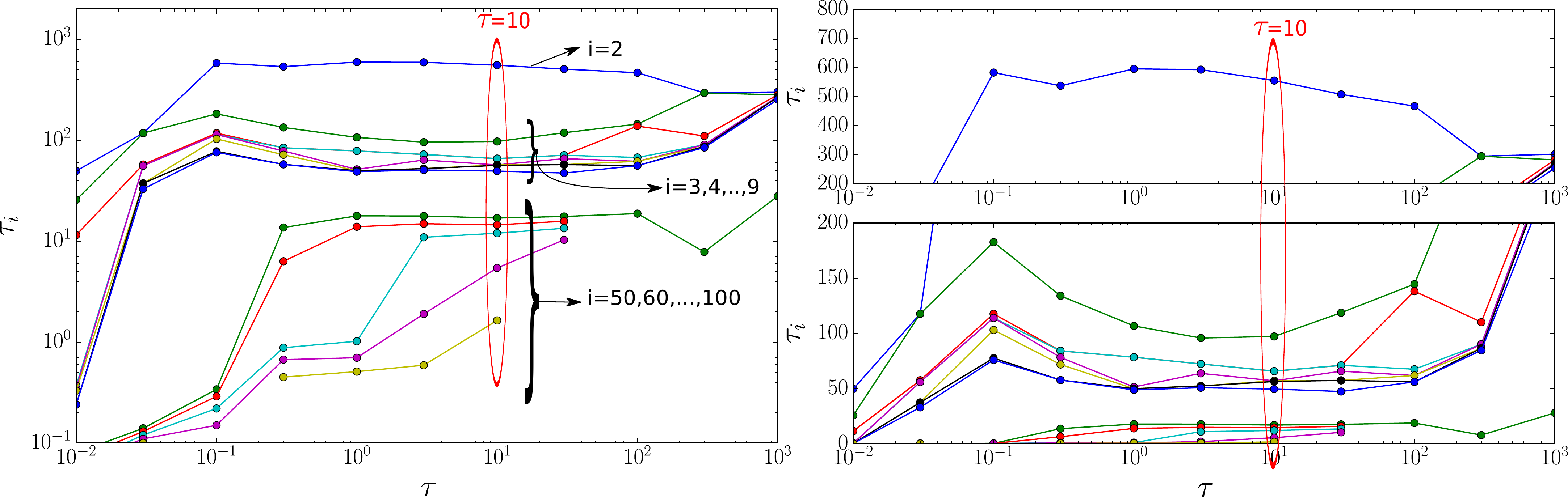}
\caption{
Timescales $\tau_i=1/\log(\lambda_i)$ for a couple of representative $i$ values (as labeled), as a function of the choice of lagtime $\tau$,
obtained from the TM of the clustering. 
Left: doubly logarithmic scale; Right: linear scale in $\tau_i$.
There is a clear plateau for all time scales $i \leq 60$ in the range $\tau\in [0.5, 50]$.
Right: same plot in semi-log coordinates, to show the details of fluctuations around the chosen value, $\tau=10$.
\label{app:tau2VSlagtime}
}
\end{figure*}

In this perspective, we start by computing the first non-trivial time scales $\tau_i, i>1$ (since $\tau_1=\infty$), for various lagtimes $\tau$.
The definition of $\tau_i$ is $-\tau/\log(\lambda_i)$.
We neglect sample-to-sample variations in this first study, as they are relatively small for $\tau_i, i<100$, compared to the dependence in $\tau$.
From Figure \ref{app:tau2VSlagtime} we see that the range of acceptable lagtimes is contained in $1-100$.
In what follows, we will use $\tau=10$ unless explicitly stated, since this value lies in the center of the range.

\begin{figure*}[h!]
\includegraphics[width=0.49\textwidth]{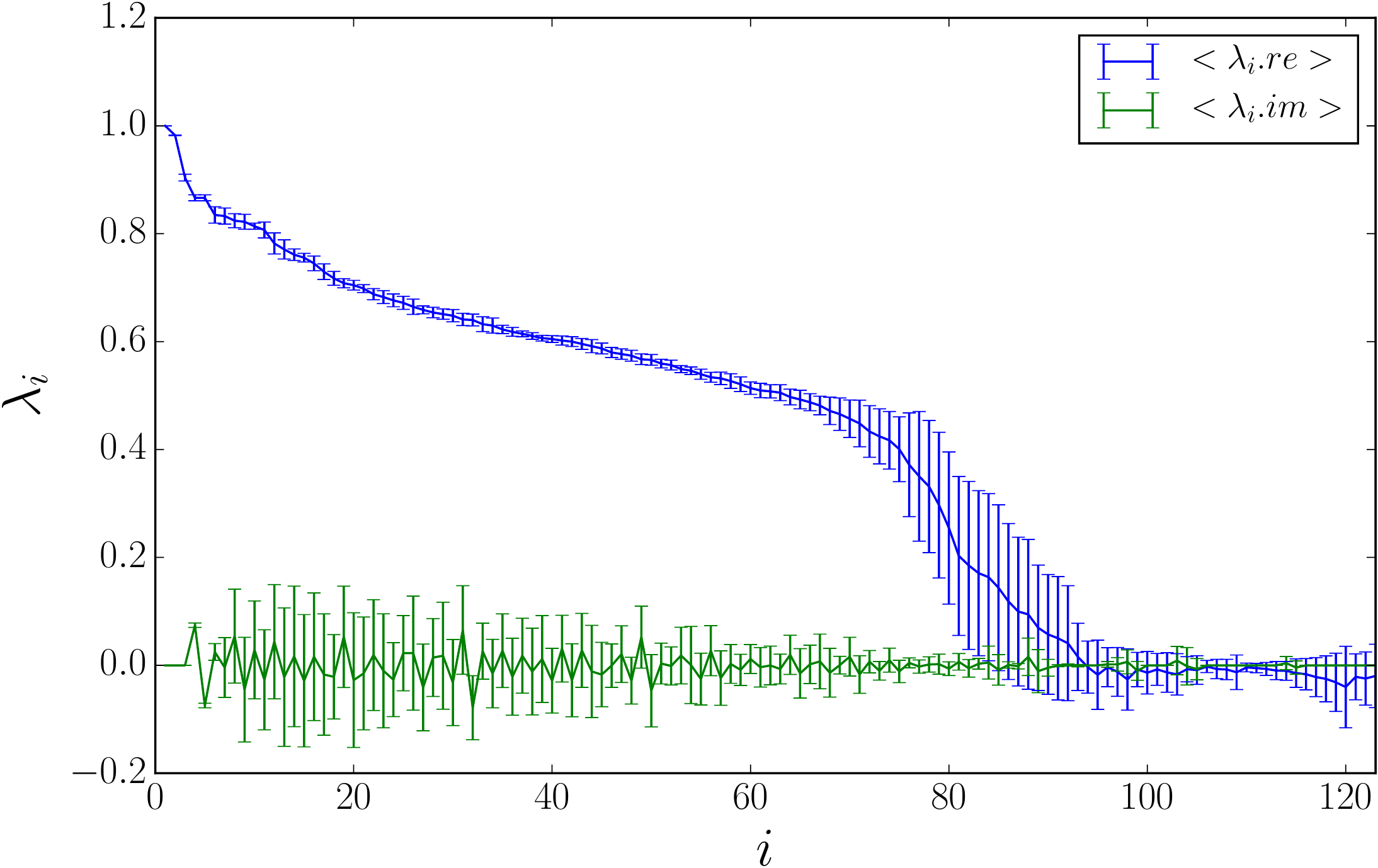}	
\includegraphics[width=0.49\textwidth]{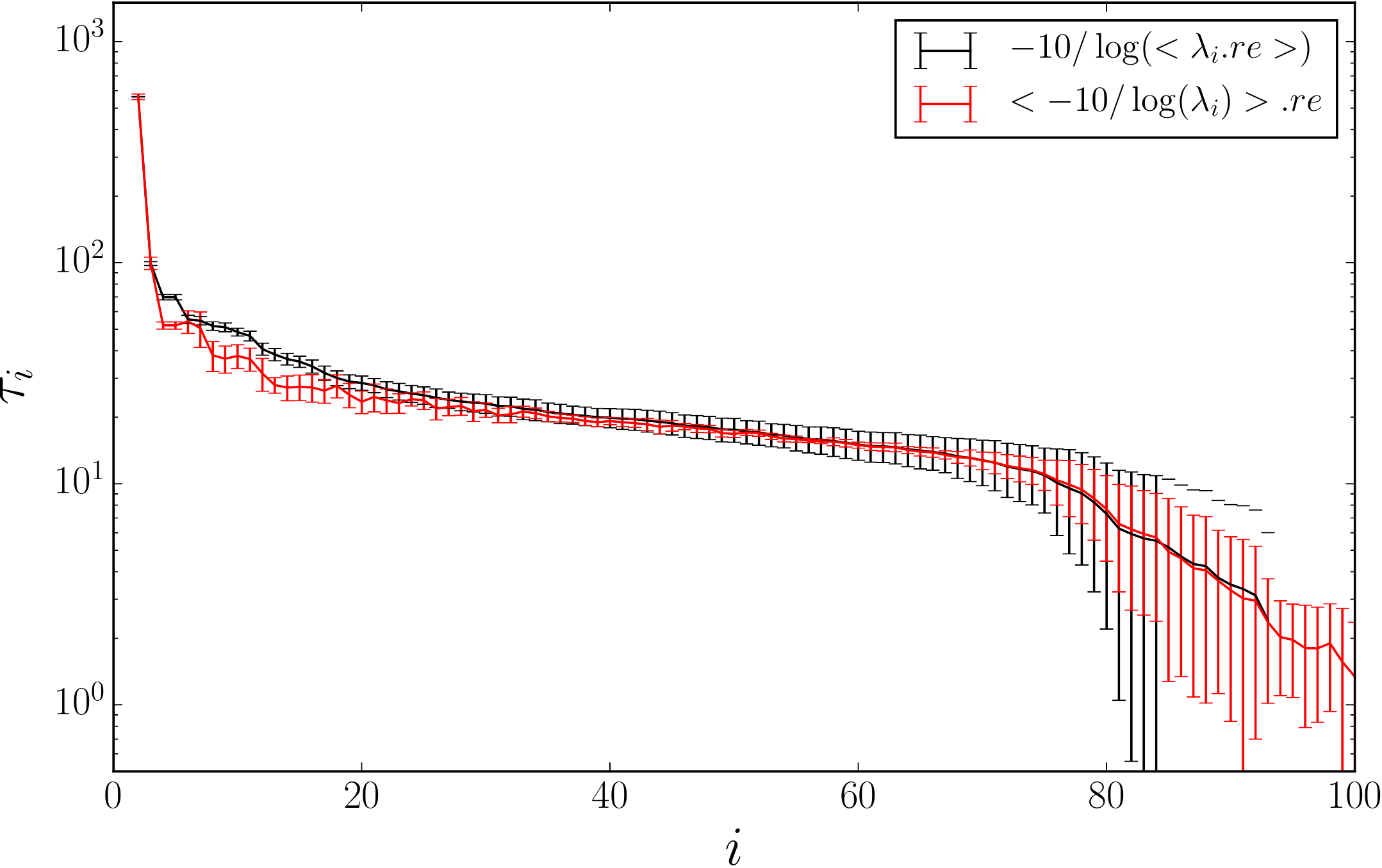}
\caption{
Full eigenvalue spectra (left) and 
timescale distribution (right) depending on the lagtime, using lagtimes $\tau=10$.
They are fairly stable for $\tau \approx 10$.
The error bars show the standard deviation.
\label{app:averages}
}
\end{figure*}

Because the sample size $N=10^5$ is still too small for the phase space to be uniformly explored, the spectra fluctuate between the sets of $N$ configurations.
To assess the reliability of these spectra, we average across 10 sets of $N$ configurations and compute the standard deviation of the results (see figure \ref{app:averages}).
Note that figure 2 of the main text was obtained using this average.

\section{2 PCCA+}

\subsection{2.1 Short presentation}

The key idea underlying standard PCCA \cite{Deuflhard2000} is that the TM relative to a group of $\tilde{n}_c$ disconnected (i.e. separate, independent) Markov chains is block-diagonal:
 the Perron eigenvalue $\lambda =1$ has degeneracy $\tilde{n}_c$ and the first $\tilde{n}_c$ (right) eigenvectors $\chi_{\alpha}$ map to membership functions $\xi_{\alpha}$ that are indicator functions, $(\xi_{\alpha})_j = \{0, 1\}$ ($1$ for all the states $j$ that belong to the Markov chain $\tilde c_{\alpha}$ and $0$ elsewhere). 
These $\xi_{\alpha}$'s are found by following the sign structure of the $\chi_{\alpha}$'s. 

A more advanced version, the Robust Perron Cluster Cluster Analysis 
(PCCA+)~\cite{Deuflhard2004, Weber2005} 
makes use of fuzzy sets, i.e. membership functions become probabilities: $\sum_j (\xi_{\alpha})_j =1$ ($(\xi_{\alpha})_j \geq 0$). 
To find these $\xi_i$'s, one has to optimize a regular matrix $\mathcal{A}$ linking the $\tilde{n}_c$ membership vectors $\xi_{\alpha}$ to the $n_c$ eigenvectors  $\chi_i$: $\widehat \xi = \mathcal{A} \widehat \chi$ ($\widehat \chi$ denotes the matrix of all eigenvectors, $\widehat \xi$ that of all membership vectors).
The microstate $j$ is then assigned to the cluster ${\text{argmax}_{\alpha}}[(\xi_{\alpha})_j]$, and we obtain the TM after re-clustering, $\widetilde{\Pi}_{\alpha\beta}$.
We must optimize $\mathcal{A}$ with respect to some cost function, usually $J_1=\sum_j {\text{max}_{\alpha}}[(\xi_{\alpha})_j]$ or $J_2=Tr(\mathcal{A}) =  \sum_\alpha \widetilde{\Pi}_{\alpha\alpha}$,
in the space of possible matrices $\mathcal{A}$.
Intuitively, the quality functions $J_1$ is the sum over the $\tilde{n}_c$ macrostates of the assignation probability of the microstate $c_\alpha$ that is assigned with best confidence, i.e.~it makes sure that each macrostate has at least one microstate that is assigned to it with large probability.
Conversely $J_2$ measures the metastability of each final macrostate: it is the sum of the weights of self-links in the final $\tilde{n}_c\times \tilde{n}_c $ graph (the trace of the final $\tilde{n}_c\times \tilde{n}_c$ transition matrix).
We optimized $J_2$, but results do not significantly change if we optimize $J_1$.

\begin{figure*}[h!]
\includegraphics[width=0.49\textwidth]{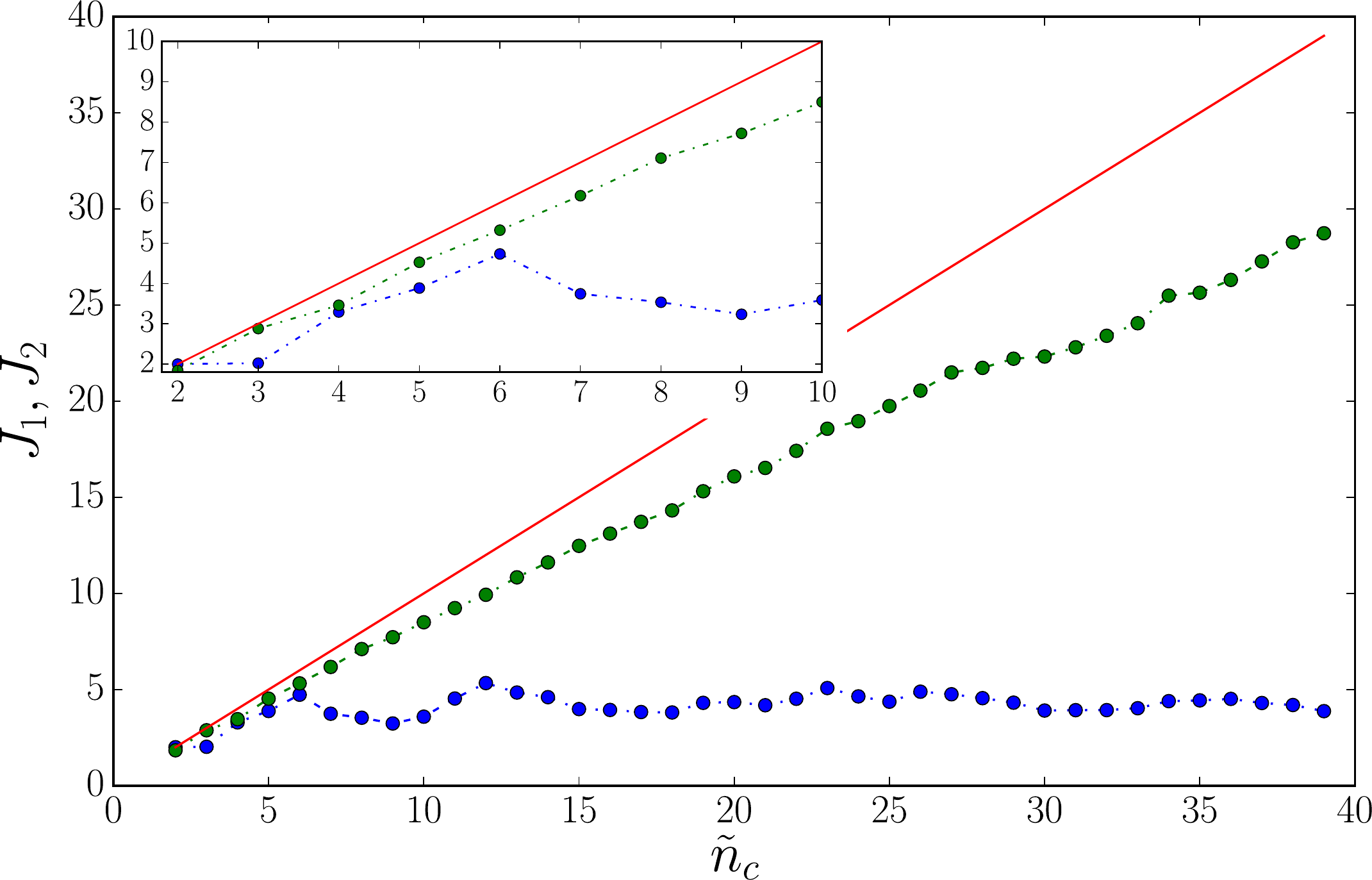}
\includegraphics[width=0.49\textwidth]{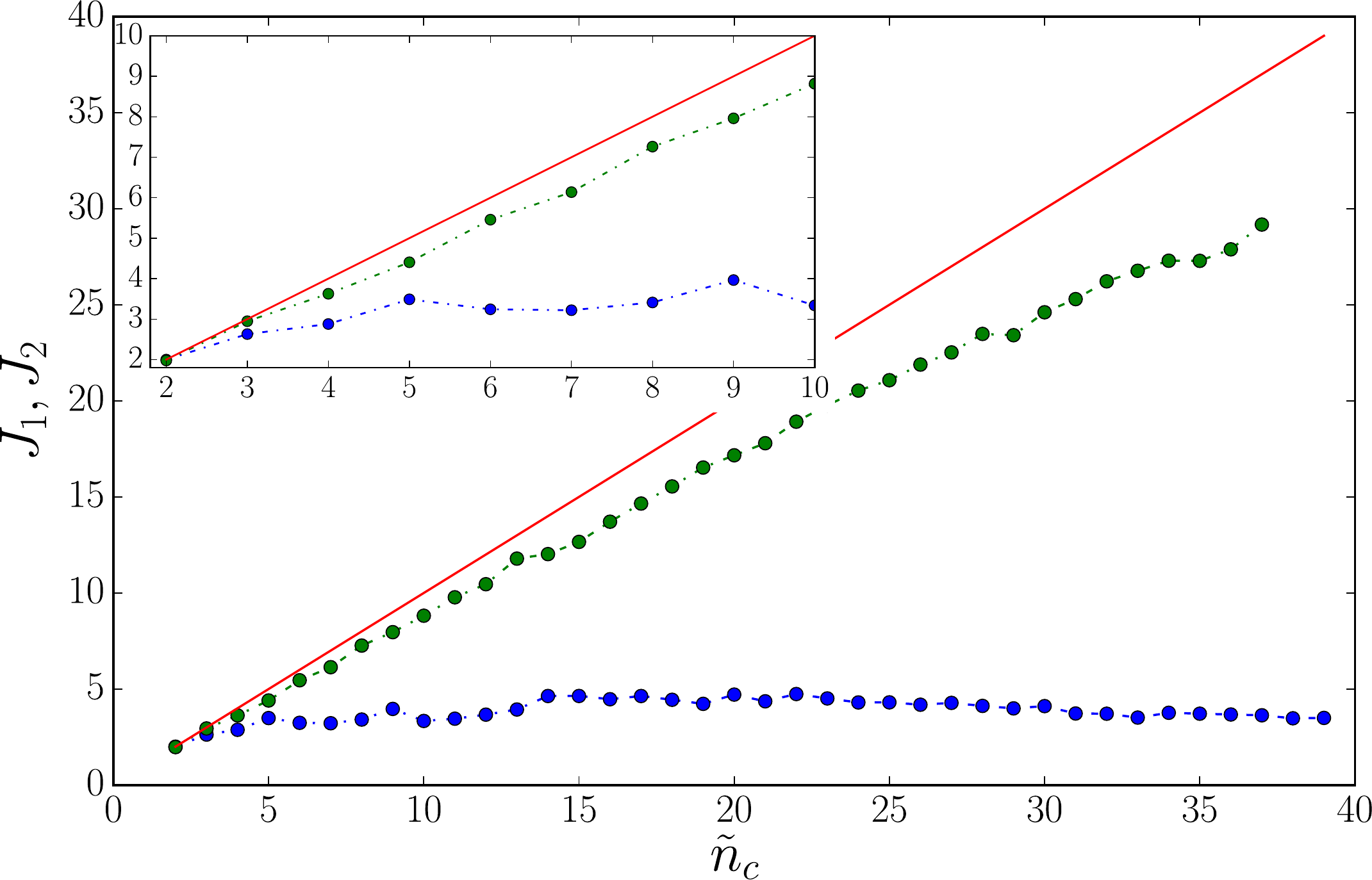}
\caption{
Cost functions $J_1$ and $J_2$ as a function of the number of macrostates $\tilde{n}_c$ in which we ask the clusters to be re-grouped.
Both are bounded by $\tilde{n}_c$ (red line).
The PCCA+ is applied on the clustering for two lengths of the chain, $L=10$ (left) and $L=15$ (right).
We note that $J_2$ does not depend strongly on $\tilde{n}_c$ (it increases almost as its upper bound).
$J_1$ has relative maxima in $6$ (for $L=10$) and 6 or 5 (for $L=15$).
We pick $\tilde{n}_c=6$ macrostates as being the largest $\tilde{n}_c$ that still optimizes $J_1$. 
The lagtime used is $\tau=10$. 
\label{app:J1J2}
}
\end{figure*}

Applying PCCA+ on a couple of clustering of $N=10^5$ configurations, each with $n_c \sim 100$ clusters, we identify $\tilde{n}_c=6$ as the optimal number of macrostates for $L=10$, and $\tilde{n}_c=5$ or $6$ for $L=15$: see Fig.~\ref{app:J1J2}.

Note that to give a decent weight to the center of mass, we define the distance between configurations at times $s$ and $t$ as:
\begin{align*}
d_{st}=\left(x^s_{\rm CM}-x^t_{\rm CM}\right)_{\text{mod }2}^2+
\frac{10}{L}\sum_{l=1}^{L-1}\left(b^s_l-b^t_l\right)^2,
\label{Eq:dist_dB}
\end{align*}
which is consistent with the main text definition (for which we had $L=10$).

Note that the original PCCA+ algorithm is intended to be used on equilibrium problems, where the TM is symmetric. Here it is not the case, and we symmetrize the matrix for simplicity. 
A more refined approach would be to use a Schur decomposition of the TM instead of a direct diagonalization, as was proposed in \cite{Roeblitz2008} and performed in detail in \cite{Conrad2015}.
As results are already satisfying here, we kept with standard PCCA+: this simply overestimates some rates of exchange, and make the rates of staying in a microstate relatively smaller. 
This illustrates the robustness of the method, since despite this approximation, we still detect well the relevant macrostates.

We implemented PCCA+ using available code (\url{https://github.com/msmbuilder/msmbuilder/})
\cite{Deuflhard2004, Kube2007, Roeblitz2008, Deuflhard2000}
 as a basis, but integrated it in our toolbox, allowing the user to estimate a correct lagtime $\tau$ and compute the optimal $\tilde n_c$ in an easy way. 
Our version of the code is available at \cite{CodeLandesPellegrini2016}, together with the codes for integrating the FK model's equations and the density peak algorithm.

\begin{figure}[h!]
\includegraphics[width=0.49\textwidth]{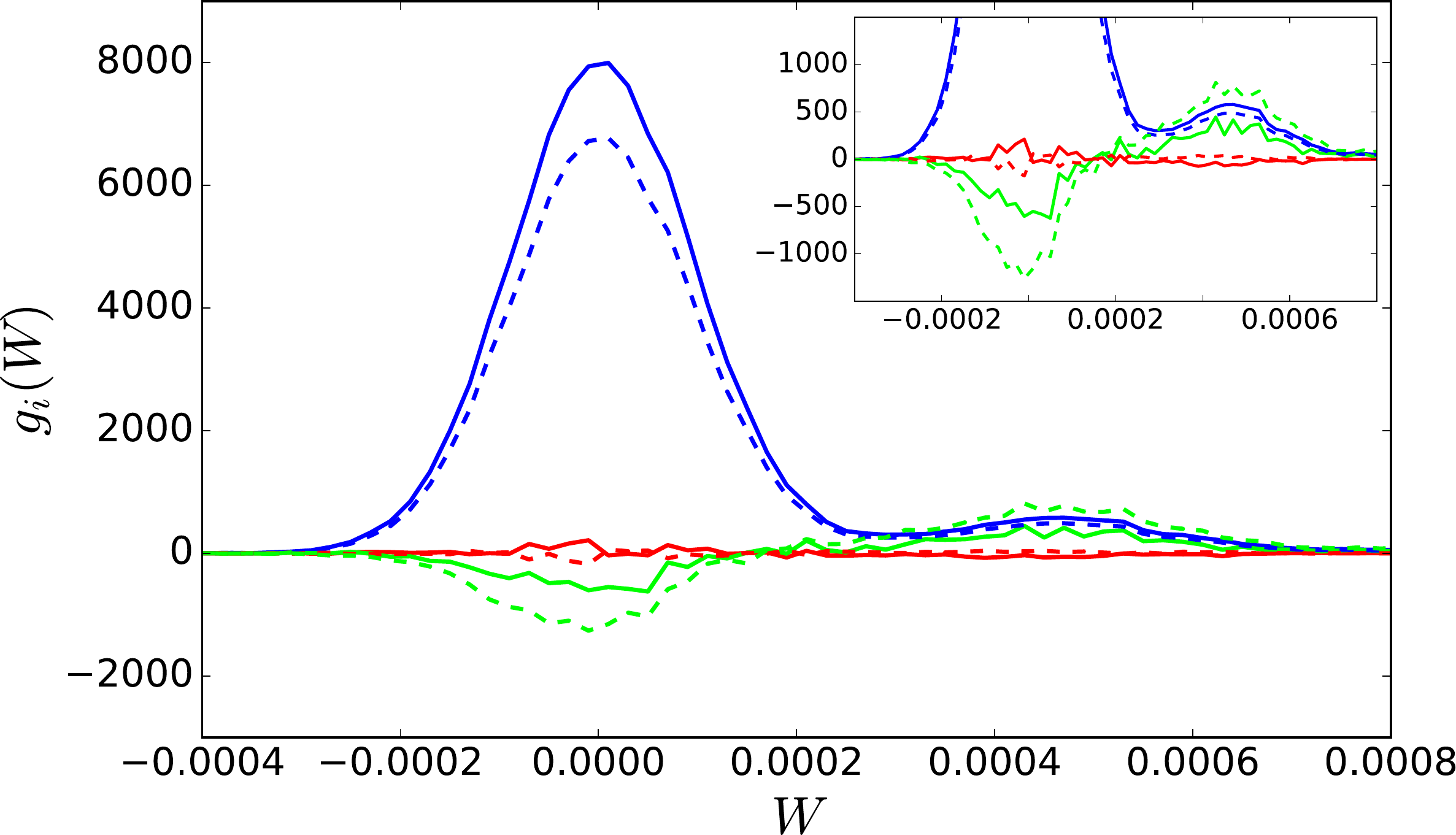}
\includegraphics[width=0.49\textwidth]{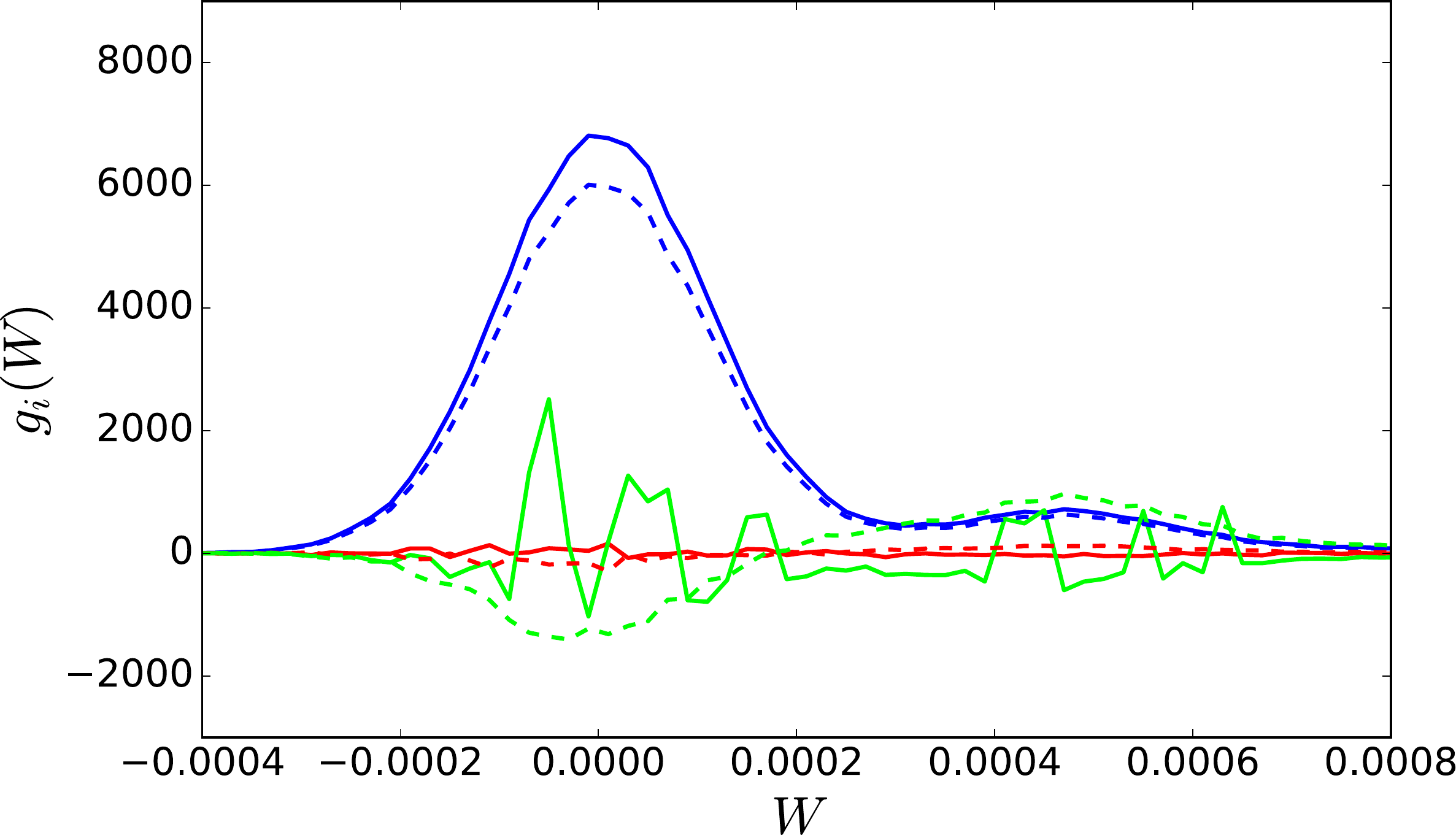}
\caption{
Comparison of work distribution $g_i(W)$  for $i=1,2,3$ (blue, red, green)
 for $L=15$ (left) and $L=20$ (right), using $\tilde{n}_c=6$ in both.
Solid lines: $n_c\sim 100 - 300$ microstate results; Dashed lines: $\tilde n_c=6$ macrostate results.
\label{fig:workN15}
}
\end{figure}

The work distribution for the chain lengths $L=15, 20$, and its excitation modes are shown in Fig. \ref{fig:workN15}.  
They are interpreted in the same way as those of Fig. 4 of the main text.
Note the robustness of the $\tilde n_c=6$ macrostate description relative to the variable $L$: the first $3$ modes are basically invariant under changes in system size.

\subsection{2.2 Further lagtime assessment}

\begin{figure}[h!]
\centering
\includegraphics[width=0.4\textwidth]{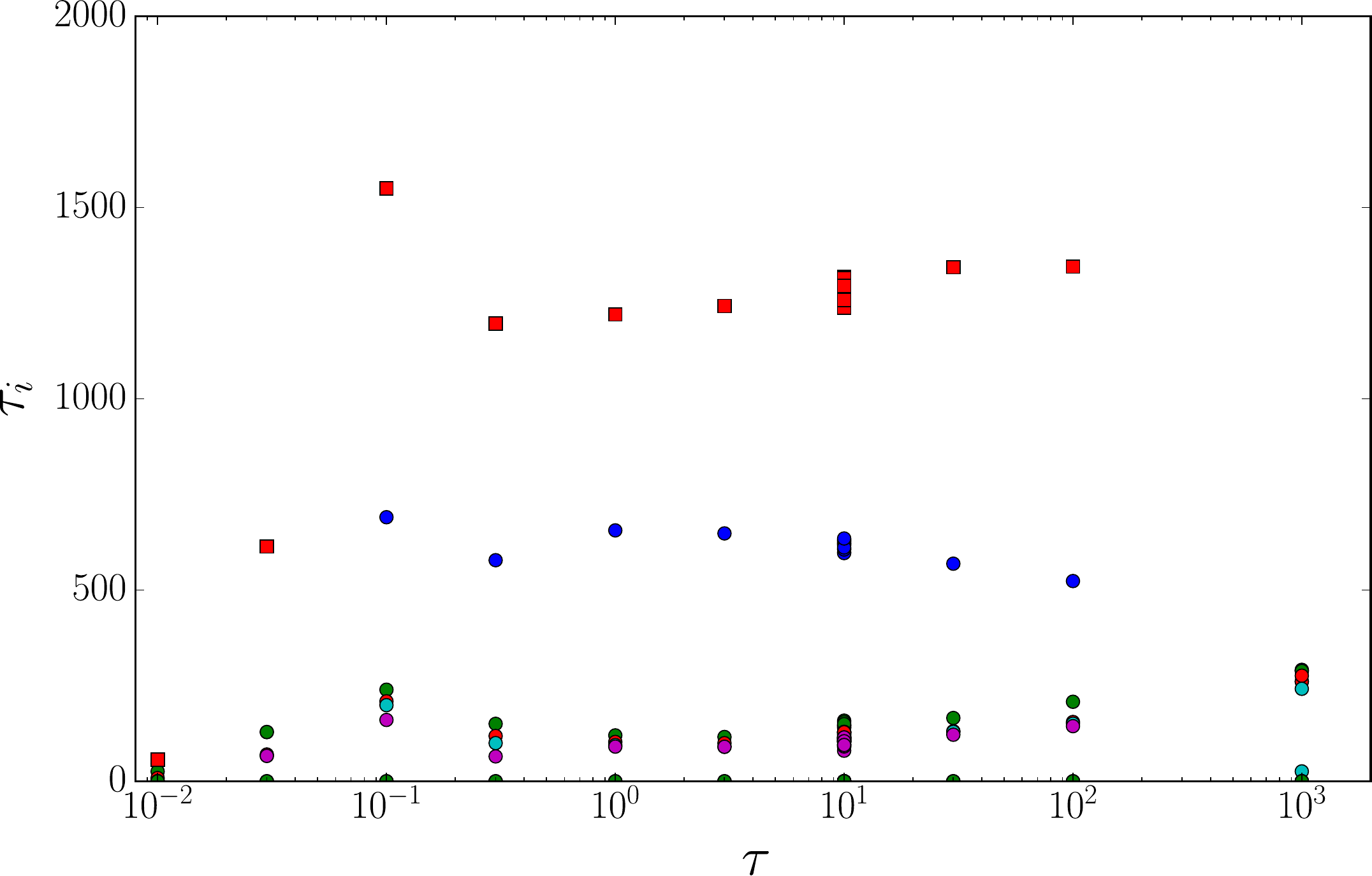}
\caption{
Time scales after lumping by PCCA into $\tilde{n}_c=6$ macrostates. 
Red 	squares show the lifetime of staying in the most probable of the $\tilde{n}_c$ states.
For the lagtime $\tau=10$ we have 10 independent realizations (left), which show little fluctuations. (for a chain of $L=10$ atoms).
\label{lumpedTimes_nn}
}
\end{figure}

One can further assess the quality of the lagtime by checking the dependence of the five timescales (since $\tilde{n}_c=6$) of the lumped model, obtained after applying PCCA+ to the clustering, together with the lifetime of staying in the most probable state (i.e. the relaxed chain, which can have either an odd or even center of mass position).
We see in Figure \ref{lumpedTimes_nn} that indeed the variations in all these timescales are small.
Our choice $\tilde{n}_c=6$ is confirmed by the study of lagtime dependence: for larger $\tilde{n}_c$, the plateaus quickly deteriorate (not shown).

\subsection{2.3 Further characterization of the lumping obtained by PCCA+}

The Fig. 3 of the main text does not show all the clusters assigned by the PCCA+, as there are too many.
To give a more complete view of how the PCCA+ lumps the clusters we found, we provide a complete description of these states in Fig. \ref{app:nnCountIntegral}.
In particular, we define the rescaled chain length (shifted by $-L$), which thus takes values between $-1$ and $1$.
We also define a quantity $k$, or KinkCount, which is a proxy for the number of defects, $k=\sum_{l=1}^{L-1}(b_{l})^{2}$, which was identified as the main collective variable.
The position of the center of mass $x_{\rm CM}$ was defined in the text.
We also show the assignation probability of each microstate $j$ to the macrostate $\alpha$ it was affected to, i.e.~the value of $p = {\text{max}_{\alpha}}[(\xi_{\alpha})_j]$. A high value indicates that the assignation was done with high confidence, while a low value means that this assignation is not robust.
The weight $w=100  P_j^{ss}$ represents the steady state probability to be in the microstate $j$ (we can take this probability according to the raw data statistics or according to $\Pi_{\alpha\beta}$, in our case it is the same, because the clustering works very well). 
These weights $w$ sum to $100$, for clarity's sake.

\begin{figure*}[h!]
\centering
\includegraphics[width=0.32\textwidth]{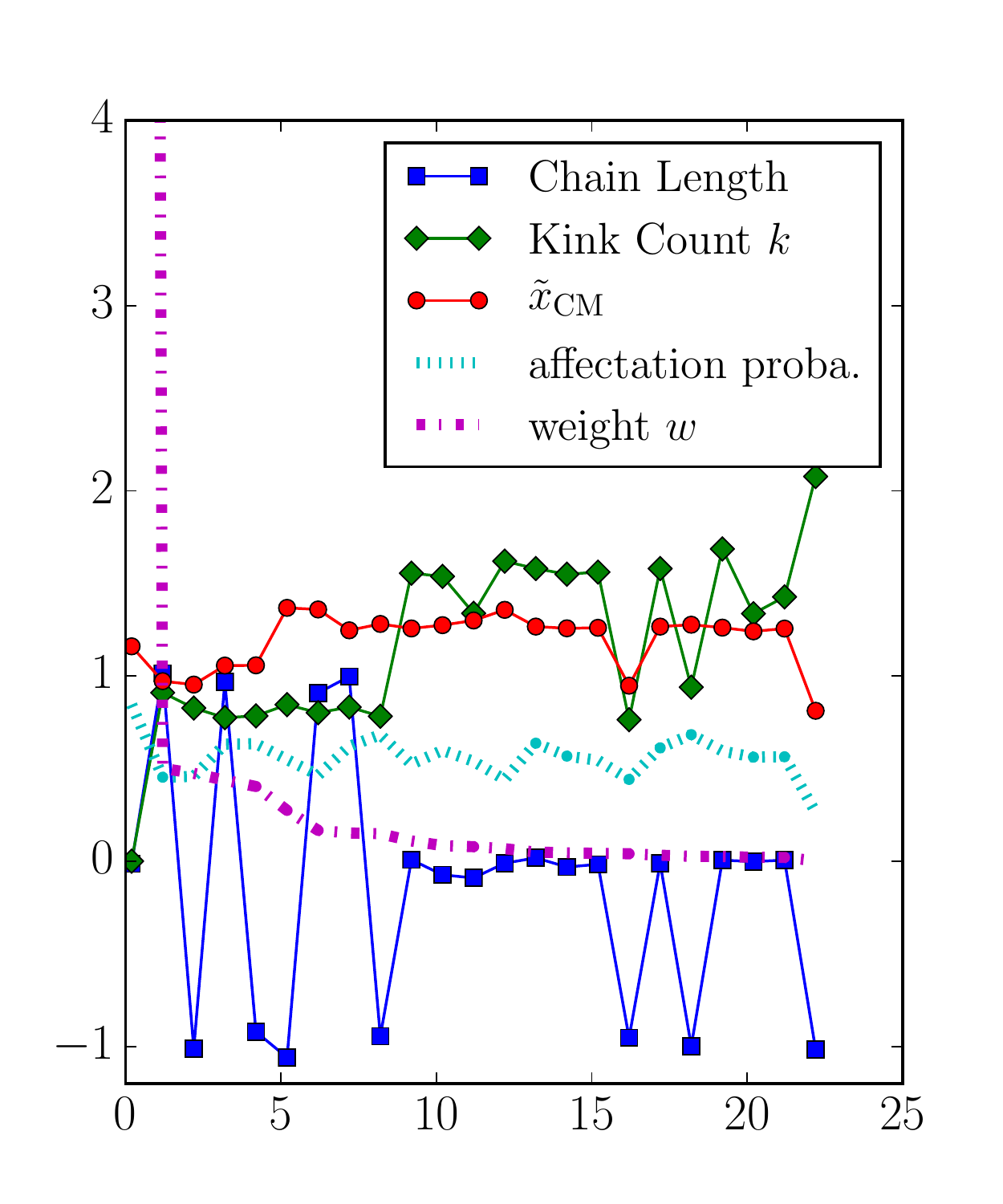}	
\includegraphics[width=0.32\textwidth]{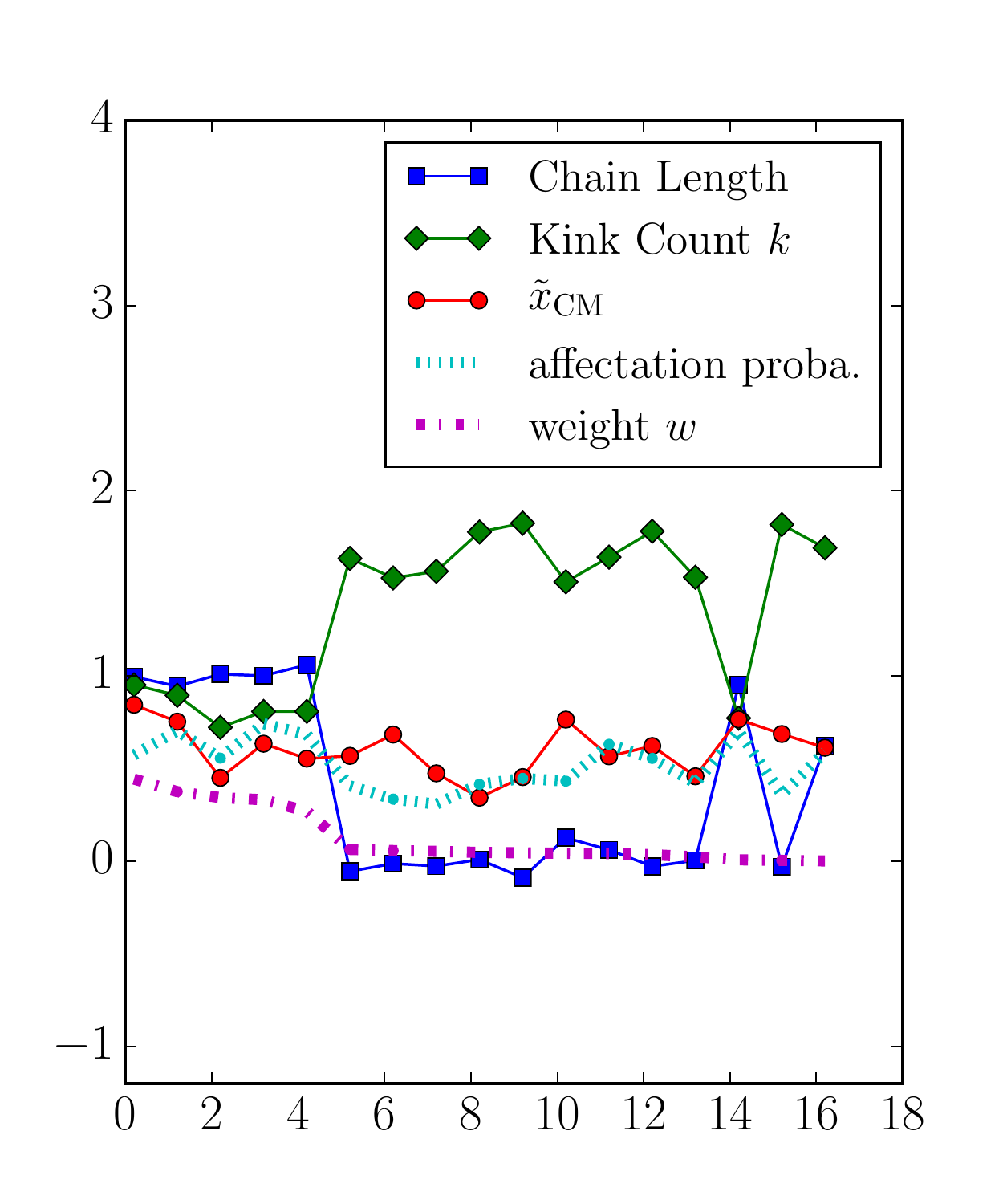}
\includegraphics[width=0.32\textwidth]{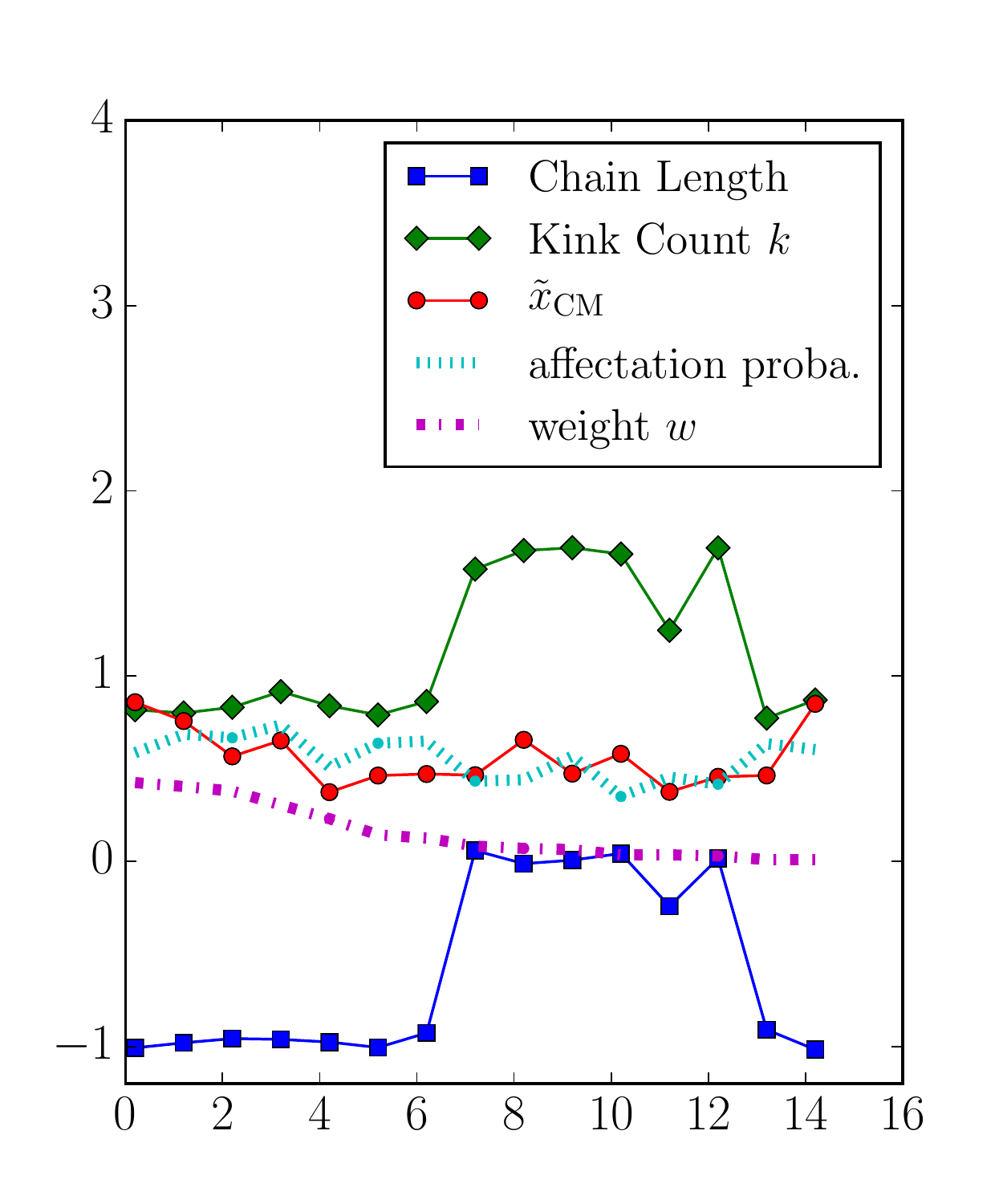}
\caption{
Each panel corresponds to a group: from left to right, $g=1,2,3$ (following the main text labels).
In each group, we show the value of the chain length for each microstate (shifted by $-L$), the  kinkCount measure, the center of mass position $\tilde{x}_{{\rm CM}}$ (rescaled and modulo 2), the assignation probability provided by the PCCA+ method, and the microstate's global weight $w$ (summing them all gives 100). There are 121 microstates, before lumping.
\label{app:nnCountIntegral}
}
\end{figure*}
\begin{figure*}[h!]
\centering
\includegraphics[width=0.32\textwidth]{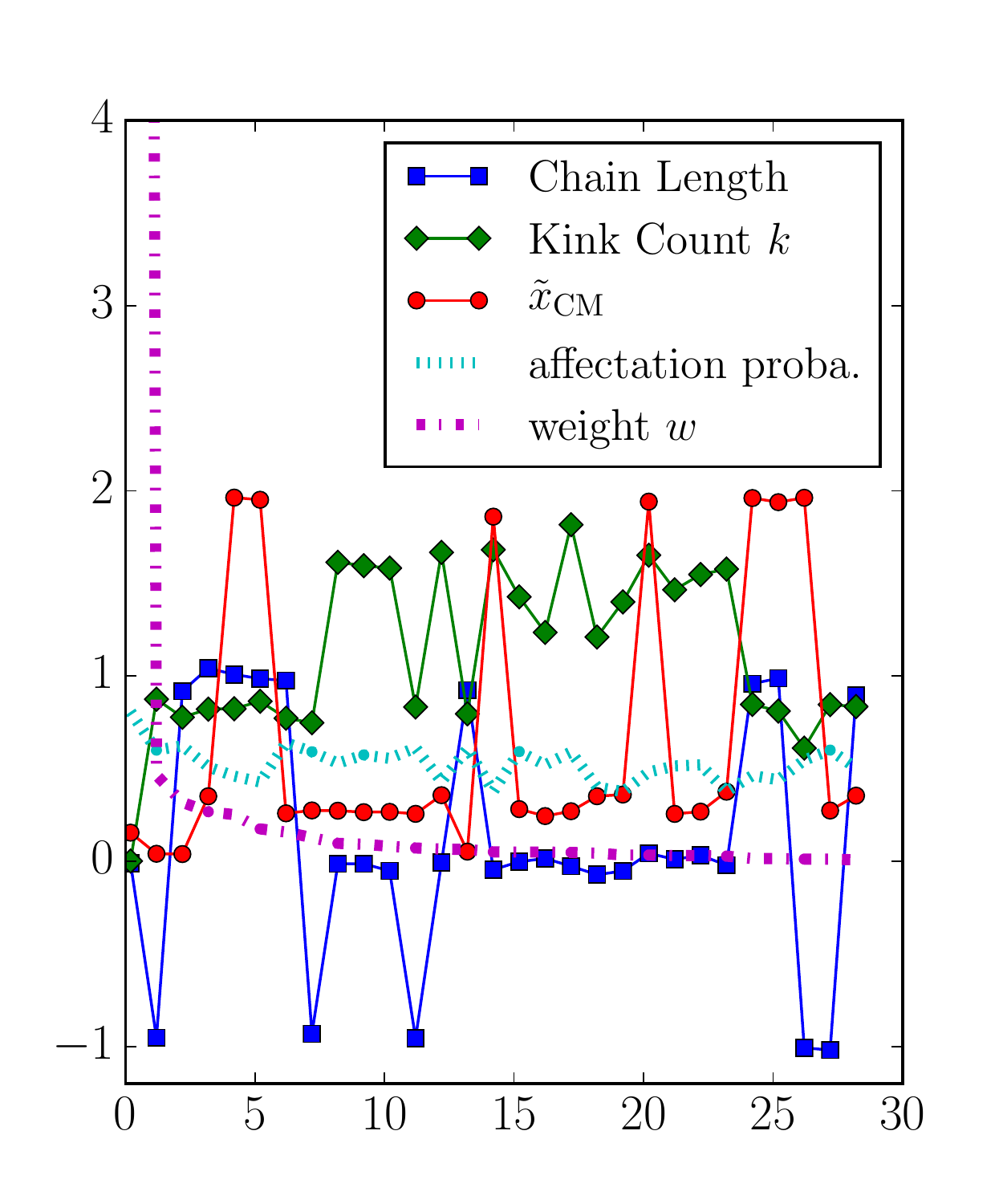} 
\includegraphics[width=0.32\textwidth]{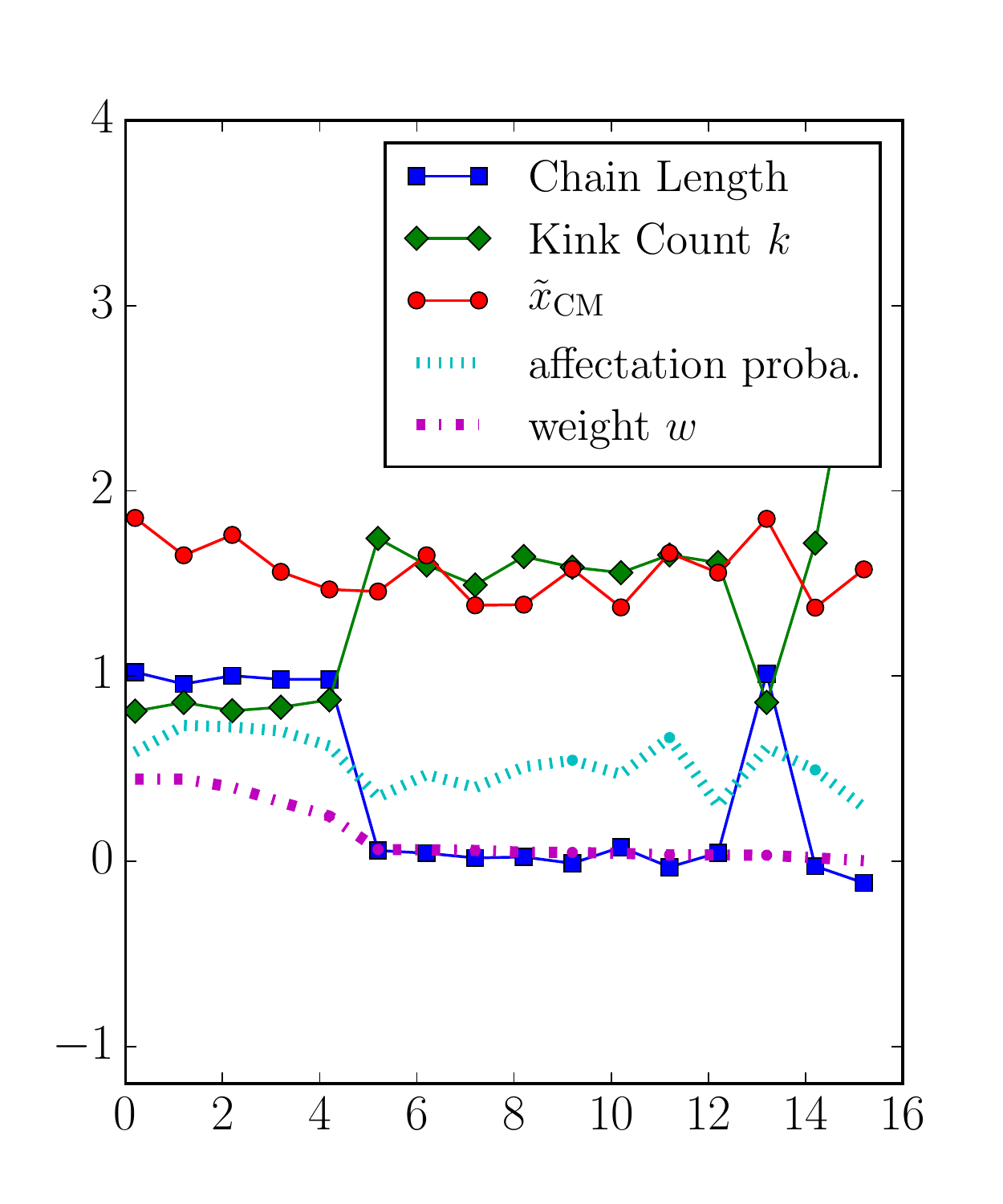}
\includegraphics[width=0.32\textwidth]{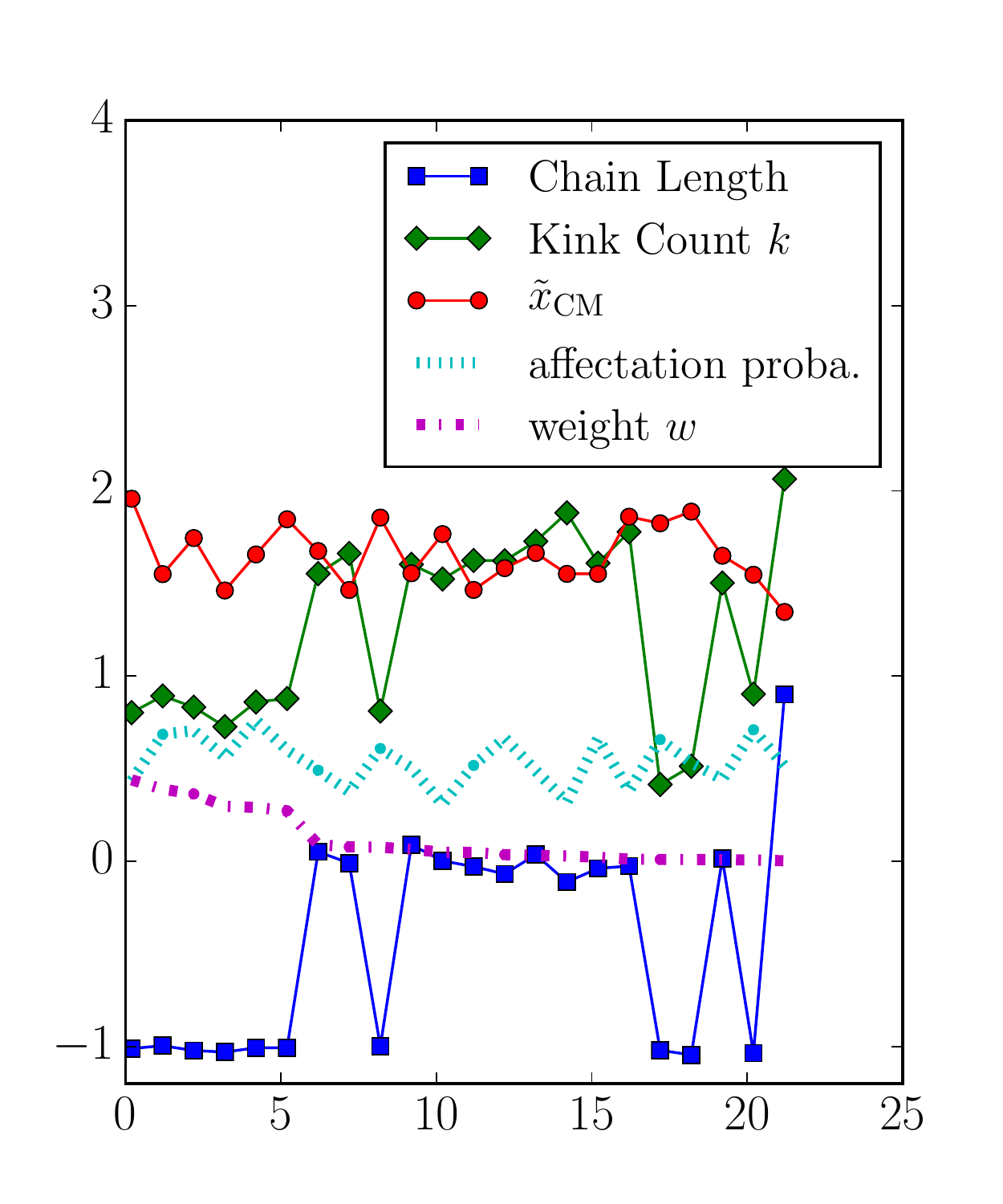}
\caption{
Continuation of Fig. \ref{app:nnCountIntegral} but for the lumps with the other parity (other value of the center of mass): here the groups are $g=4,5,6$ (left to right).
Note that these 3 panels are extremely similar to the three above, as expected, except of course for the center of mass position, which is around $0.15$ (or $2$ because of periodicity) in one case and around $1.15$ in the other.
\label{app:nnCountIntegral2}
}
\end{figure*}

\end{document}